\documentclass[superscriptaddress,twocolumn,secnumarabic,amssymb, nobibnotes, aps, prd,showpacs]{revtex4-1}

\usepackage{graphicx}
\usepackage{amsmath}

\begin{document}

\title{Structural characterization of PrVO$_3$ epitaxial thin films}%

\author{O. Copie}%
%\email{olivier.copie@ensicaen.fr}
\altaffiliation{Present address: Laboratoire SPMS, CNRS-Ecole Centrale Paris (UMR 8580), Grande voie des vignes, 92295 Ch\^atenay-Malabry Cedex, France}
\affiliation{Laboratoire CRISMAT, CNRS-ENSICAEN and Universit\'e de Caen Basse Normandie (UMR 6508), 6 boulevard du Mar\'echal Juin, 14050 Caen Cedex, France}

\author{H. Rotella}
\altaffiliation{Present address: National University of Singapore (NUS), 117542 Singapore}
\affiliation{Laboratoire CRISMAT, CNRS-ENSICAEN and Universit\'e de Caen Basse Normandie (UMR 6508), 6 boulevard du Mar\'echal Juin, 14050 Caen Cedex, France}

\author{P. Boullay}
\affiliation{Laboratoire CRISMAT, CNRS-ENSICAEN and Universit\'e de Caen Basse Normandie (UMR 6508), 6 boulevard du Mar\'echal Juin, 14050 Caen Cedex, France}

\author{M. Morales}
\affiliation{CIMAP, CNRS-ENSICAEN and Universit\'e de Caen Basse Normandie (UMR 6552), 6 boulevard du Mar\'echal Juin, 14050 Caen Cedex, France}

\author{A. Pautrat}
\affiliation{Laboratoire CRISMAT, CNRS-ENSICAEN and Universit\'e de Caen Basse Normandie (UMR 6508), 6 boulevard du Mar\'echal Juin, 14050 Caen Cedex, France}

\author{A. David}
\affiliation{Laboratoire CRISMAT, CNRS-ENSICAEN and Universit\'e de Caen Basse Normandie (UMR 6508), 6 boulevard du Mar\'echal Juin, 14050 Caen Cedex, France}

\author{B. Mercey}
\affiliation{Laboratoire CRISMAT, CNRS-ENSICAEN and Universit\'e de Caen Basse Normandie (UMR 6508), 6 boulevard du Mar\'echal Juin, 14050 Caen Cedex, France}

\author{D. Pravarthana}
\affiliation{Laboratoire CRISMAT, CNRS-ENSICAEN and Universit\'e de Caen Basse Normandie (UMR 6508), 6 boulevard du Mar\'echal Juin, 14050 Caen Cedex, France}

\author{I. C. Infante}
\affiliation{Laboratoire SPMS, CNRS-Ecole Centrale Paris (UMR 8580), Grande voie des vignes, 92295 Ch\^atenay-Malabry Cedex, France}

\author{P.-E. Janolin}
\affiliation{Laboratoire SPMS, CNRS-Ecole Centrale Paris (UMR 8580), Grande voie des vignes, 92295 Ch\^atenay-Malabry Cedex, France}

\author{W. Prellier}
%\email{wilfrid.prellier@ensicaen.fr}
\affiliation{Laboratoire CRISMAT, CNRS-ENSICAEN and Universit\'e de Caen Basse Normandie (UMR 6508), 6 boulevard du Mar\'echal Juin, 14050 Caen Cedex, France}

%\date{\today}%

\begin{abstract}
Rare earth perovskite oxides constitute a wide family of materials presenting functional properties strongly coupled to their crystalline structure. Here, we report on the experimental results on epitaxial PrVO$_3$ deposited on SrTiO$_3$ single crystal substrates by pulsed laser deposition. By combining advanced structural characterization tools, we have observed that the PVO unrelaxed film structure grown on STO, is characterized by two kinds of oriented domains whose epitaxial relations are: (i) PrVO$_3[110]_o\|$SrTiO$_3[001]_c$ and PrVO$_3[001]_o\|$SrTiO$_3[100]_c$, (ii) PrVO$_3[110]_o\|$SrTiO$_3[001]_c$ and PrVO$_3[001]_o\|$SrTiO$_3[010]_c$. We have also measured reciprocal space maps. From these results, we have determined that the PVO film epitaxy on STO imposes a lowering of the PVO structure symmetry from orthorhombic ($Pbnm$) to monoclinic ($P2_1/m$). We show, the nominal strain induced by the substrate being constant, that the obtained film structure depends on both growth oxygen and temperature. Thus, by finely controlling the deposition conditions, we could tune the strain experienced by PrVO$_3$ thin film. These results show an alternative to substrate mismatch as a path to control the strain and structure of PVO films 
\end{abstract}


\maketitle

\section{Introduction}
Transition metal oxides (TMO) with a perovskite structure display a wealth of functional properties that are often governed by the interplay of the charge, spin, orbital and lattice degrees of freedom~\cite{Imada1998,Tokura2000}.The great potential of these materials as the future basis of the emerging oxide electronics~\cite{Bibes2011} has stimulated an increasing attention, particularly when TMO gather more than one property such as in multiferroics~\cite{Chu2008}. Particularly relevant for their future integration in microelectronic devices, the research on TMO thin films has strongly benefitted from the development of physical vapor deposition techniques~\cite{Schlom2008}. The origin of their multifunctional character being entangled to crystalline structure, a large landscape of tunabilities through film growth is possible for understanding the coupling mechanisms in these systems. The role of oxide epitaxy for tuning the functionality or for combining very different properties~\cite{Chakhalian2006} has accompanied the development of the growth techniques. In this context, lattice mismatch control has led to impressive results on epitaxially driven ferroelectric~\cite{Haeni2004}, ferromagnetic~\cite{Rubi2009,Marti2009,White2013} or multiferroic~\cite{Lee2010} properties in thin films, such properties being originally absent in the bulk form.

The relation between spin order (SO) and orbital order (OO), coupled to lattice distortions in TMO, is highly studied in the prototypical perovskite orthovanadates. Indeed, different spin and orbital ordered phases are stabilized with temperature, arising from the coupling of spin, orbital and lattice degrees of freedom~\cite{Miyasaka2003,Zhou2006}. In RVO$_3$ (R = rare-earth or yttrium), a G-type orbital ordering (G-OO) takes place at the temperature T$_{\rm{OO}}$ concomitant with a structural phase transition (orthorhombic to monoclinic).  Below T$_{\rm{OO}}$, superexchange interactions drive the system into a C-type antiferromagnetic spin order (C-SO) below the so-called the temperature T$_{\rm{SO}}$. For small R-ions (R = Dy-Lu), a transition towards C-OO and G-SO occurs at lower temperature~\cite{Miyasaka2003}. Remarkably, it has be shown that T$_{\rm{OO}}$ can be tuned either by chemical pressure, by decreasing the R-site ionic radius~\cite{Miyasaka2003}, by a partial substitution of R-site with alkaline earth~\cite{Fujioka2005} or physically by applying hydrostatic pressure~\cite{Zhou2007,Bizen2008}, attesting a strong orbital-lattice coupling. Then, the strained structure further affects the onset of the magnetic transitions~\cite{Zhou2007,Bizen2012}. Reversely, the magnetic-field induced G- to C-SO transition (together with C- to G-OO transistion), reported in DyVO$_3$ illustrates also the intimate spin-orbital interaction~\cite{Miyasaka2007} that could suggest a possible multiferroic behavior~\cite{Zhang}. We have recently reported the optimized thin film growth of a member of the RVO$_3$ series, PrVO$_3$ (PVO), by pulsed laser deposition\cite{Copie2013}. We evidenced a semiconducting character in good agreement with bulk characteristics. However, our magnetic characterization revealed a significant ferromagnetic-like behavior at low temperature likely induced by uncompensated spins of a canted antiferromagnetic state~\cite{Mahajan1992,Onoda1996,Ren1998}. Furthermore, we observed that T$_{\rm{SO}}$ is $\sim60$~K lower than the bulk spin ordering temperature ($\sim135$~K). Our preliminary structural measurements reveals a larger pseudocubic PVO unit cell volume. In view of these results, we have attributed the T$_{\rm{SO}}$ lowering to a weaker neighboring exchange interactions when PVO is grown in thin film.

In order to rationalize the influence of the strained lattice on the magnetic behavior, we present here a detailed study of the PVO films by combining advanced structural characterization tools. We have investigated the structural properties of various PVO thin films and we have observed that a fine tuning of the deposition conditions yields a control of the PVO unit cell distortion. Indeed, we show that tunable and sizable strain state is obtained and depends on both growth pressure and temperature, the nominal epitaxial strain imposed by the substrate being constant.
\begin{figure}[tp]
 \includegraphics[scale=1]{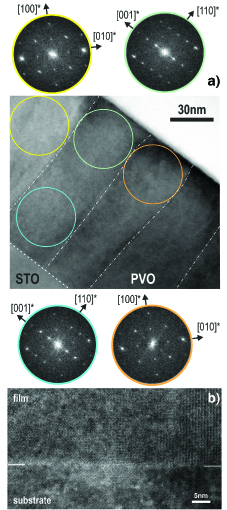}
    \caption{(Colour online)  (a) Cross-sectionnal HRTEM image of a PVO film (upper) on STO (bottom) taken along the cubic [010] zone axis. Encircled patterns are the Fourier transforms (FT) of the encircled region in the HRTEM image. The existence of extra reflections in FT as well as the contrast in the HRTEM image attest that PVO adopts a structure different from a simple perovskite. Upper left and lower right FT correspond to the upper left (yellow) and right (orange) encircled region while the upper right and lower left FT correspond to the upper middle (green) and lower left (blue) encircled region. (b) Cross-section HRTEM image close to the PVO/STO interface confirming the epitaxial growth of PVO on STO. The film/substrate interface is indicated by horizontal white lines.}
    \label{Fig1}
\end{figure}
\section{EXPERIMENTAL DETAILS}
At room temperature, bulk PVO structure adopts an orthorhombic \textit{Pbnm} crystal structure (group symmetry $\#$62) with the lattice parameters $a_0=5.487$~\AA, $b_0=5.564$~\AA~and $c_0=7.779$~\AA. The orbital and spin orderings take place at T$_{\rm{OO}}\approx 180$~K and  T$_{\rm{SO}}\approx 135$~K, respectively~\cite{Miyasaka2003,Tung2005,Sage2007}.
The PVO films were grown by pulsed laser deposition using a KrF laser ($\lambda_{\text{laser}}=248$~nm) at repetition rate of 2 Hz with a laser fluence of 2~Jcm$^{-2}$ focusing on a PrVO$_4$ polycrystalline target prepared by standard solid-state reaction. The films were deposited  on (001)-oriented SrTiO$_3$ (STO) single crystalline cubic substrates ($a_{s}=3.905$ \AA). The temperature (T$_{growth}$) and	the	pressure (P$_{growth}$) have been varied between 500$^{\circ}$C and	800$^{\circ}$C	and	between 1$\times10^{-3}$~mbar and 7.4$\times10^{-6}$~mbar, respectively. Energy dispersive spectrometry (EDS) was performed using a FEI XL30 scanning electron microscope. Sample cross-sections and planar-views were prepared and then observed using high-resolution transmission electron microscopy (HRTEM) with a FEI Tecnai G30 microscope equipped with electron diffraction. X-ray diffraction (XRD) characterizations, using monochromatic Cu K$_{\alpha1}$ radiation, were carried out with a Philips X'Pert Pro MRD diffractometer equipped with a triple axis crystal analyzer and a Seifert diffractometer to perform reciprocal space maps (RSM) and symmetric $\theta-2\theta$ measurements, respectively.
\section{RESULTS AND DISCUSSION}
\begin{figure}[tp]
  \includegraphics[scale=1]{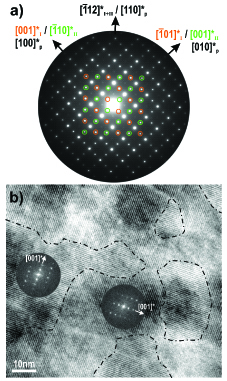}
    \caption{(Colour online) (a) Electron diffraction of a planar-view HRTEM image showed in (b). Green and orange encircled reflections correspond to the two orthorhombic domains oriented at 90$^{\circ}$ from each other. Very weak reflection intensity corresponding to the out of plane [001] orthorhombic direction is seen. (b) HRTEM plane view image of PVO take along the cubic [001] zone axis. FT indicate the different PVO orthorhombic oriented domains and the black dashed-dotted line mark the domain boundaries.}
    \label{Fig2}
\end{figure}
 \begin{figure*}[tp]
  \includegraphics[scale=1]{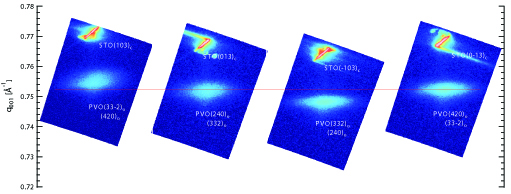}
    \caption{(Colour online) Reciprocal space maps around (from left to right) the STO (103)$_c$ and PVO (240)$_o$ reflections, STO (013)$_c$ and PVO (33$\bar{2}$)$_o$ reflections, STO ($\bar{1}$03)$_c$ and PVO (420)$_o$ reflections, STO (0$\bar{1}$3)$_c$ and PVO (33$\bar{2}$)$_o$ reflections.}
    \label{Fig3}
\end{figure*}
%
%
%
%
% FIG 1 Ð HRTEM CROSS-SECTION
%
%
%
We have first characterized the chemical composition of various PVO thin film by EDS. Therefore, we performed EDS and we found a Pr/Sr ratio close to 1:1. Within the experimental accuracy of the technique, we can rule out cation deficiency in the films.

To investigate the structural properties at the nanoscale of the PVO films, we have performed HRTEM experiments. In Fig.~\ref{Fig1} (a), we present a cross-section image of a 120~nm-thick film grown at 750$^{\circ}$C and 10$^{-5}$~mbar. These observations attest the high structural quality of PVO. The PVO/STO interface is indicated by a white dashed line. We see clearly the existence of oriented domains whose size is $\sim45$~nm with sharp boundaries (white dashed-dotted lines)~\cite{Copie2013}. Furthermore, the Fourier transform (FT) extracted from the selected regions (colored circles) exhibit extra reflections that are not related to a simple perovskite cubic lattice. Indeed, considering that bulk PVO presents a distorted perovskite structure involving the VO$_6$ octahedra rotations, this leads to a doubling of the unit cell along the $[001]_o$ direction consistent with a $Pbnm$ crystal symmetry($o$ stands for orthorhombic). Thus, we can identify the FT patterns as those from $[1\bar{1}0]_o$ and $[001]_o$ zone axes of an orthorhombic structure (Fig.\ref{Fig1}(a)) and the deduced orientations of the PVO domains correspond to two orthorhombic perovskite cells that differ by an in-plane 90$^{\circ}$ rotation. The epitaxial relation is in both cases PVO$[110]_o\|$STO$[001]_c$ ($c$ stands for cubic). We have also observed the PVO/STO interface whose HRTEM image is shown in Fig.~\ref{Fig1}(b). We see that the PVO film is perfectly strained and coherently grown onto the STO substrate so that the domain structure is preserved from the film/substrate interface to the film surface with relaxation. This is evidenced by the contrast of Fig.~\ref{Fig1}(a) and confirmed by FT of the region encircled in blue and green (Fig.~\ref{Fig1}(a)). These HRTEM results indicate that two orthorhombic domain configurations are allowed by the epitaxial growth of PVO on a cubic STO substrate. Furthermore, as seen in Fig.~\ref{Fig1}(a) that the epitaxial relations are fully preserved from the film/substrate interface to the film surface. 

Having seen that the oriented domain configuration expands on the whole film thickness and then to explore the in-plane crystalline microstructure and distribution of domaines, we have prepared planar-views a 92-nm thick PVO film grow at 600$^{\circ}$C and 8.6$\times$10$^{-6}$~mbar. The results of the HRTEM observations are displayed in Fig.~\ref{Fig2}. Electron diffraction experiments (Fig.~\ref{Fig2}(a) indicate the in-plane orientation of the PVO film. The bright reflections refer to the simple perovskite cubic lattice. The green and orange encircled extra reflections are characteristic of two possible orthorhombic domains rotated one to the other by 90$^{\circ}$ (see Fig.~\ref{Fig2}(a)). We note that very weak intensity is also observed between green and orange encircled reflections, which would correspond to the epitaxial relation PVO$[001]_o\|$STO$[001]_o$,  \textit{i.e.} the long orthorhombic axis in the growth direction. Although this orientation is allowed, it is negligible compared to the other two in-plane possibilities. Indeed, these families of PVO$[110]_o\|$STO$[001]_c$ domains are unambiguously distinguished in the HRTEM image in Fig.~\ref{Fig2}(b) with sharp domain boundaries (black dashed-dotted line). The patterns of FT performed on the different domains of Fig.~\ref{Fig2}(b) also emphasized the presence of oriented orthorhombic domains. At this stage, we observe that the PVO unrelaxed film structure grown on STO, is characterized by two kinds of oriented domains whose epitaxial relations are: (i) PVO$[110]_o\|$STO$[001]_c$ and PVO$[001]_o\|$STO$[100]_c$, (ii) PVO$[110]_o\|$STO$[001]_c$ and PVO$[001]_o\|$STO$[010]_c$. 
%
%
% FIG 2 Ð HRTEM PLANE-VIEW
%
%
\begin{table*}
\caption{\label{Table1}Bulk PVO and refined thin film lattice parameters}
	\begin{ruledtabular}
	\begin{tabular}{lllllllllllllll}
	&$a$ (\AA)& $b$ (\AA)& $c$ (\AA)& $\alpha$ ($^{\circ}$) & $\beta$  ($^{\circ}$)& $\gamma$  ($^{\circ}$)& V (\AA$^3$) &d$_{1\bar{1}0}$ (\AA)& d$_{110}$ (\AA)&  $\epsilon_{100}(\%)$ & $\epsilon_{010}$ (\%)& $\epsilon_{001}$ (\%)& $\epsilon_{1\bar{1}0}$ (\%)& $\epsilon_{110}$ (\%)\\
	bulk\footnote{from ref.~\cite{Sage2007}}&5.487& 5.564 & 7.779 & 90 & 90 & 90& 237.5 & 3.907 & 3.907 & - & - & - & - & - \\
	film&5.575(5) & 5.588(4) & 7.809(4) & 90 & 90 & 88.8(4)& 243.3 & 3.905 & 3.989(4) & $+1.6$ & $+0.4$& $+0.4$& $-0.05$&$+2.1$\\
	\end{tabular}
	\end{ruledtabular}
\end{table*}
%
% 
%
% 
%
% FIG 3 Ð XRD
%

\begin{figure*}[tp]
  \includegraphics[scale=0.8]{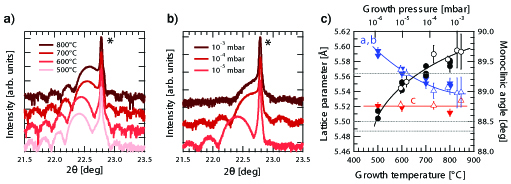}
    \caption{(Colour online) (a) and (b): $\theta/2\theta$ x-ray diffraction measurements of PVO $(110)_o$ Bragg peak close to the on $(001)_c$-oriented STO, asterisks indicate substrate $(001)_o$ peak position for a 40 nm-thick film grown at different temperatures and a 25 nm-thick film grown at different pressures, respectively. Laue fringes and indicate the good crystallinity and the PVO/STO interface quality. (c) PVO lattice parameters as a function of the growth conditions: $a$ and $b$  (blue triangle symbols), $c$ (red triangle symbols) and monoclinic angle $\gamma$ (black circle symbols) as a function of the growth temperature (bottom scale, full symbols) and as a function of the growth pressure (top scale, open symbols). Lines are guides for the eyes. Light grey dotted line represent the value of the bulk lattice parameters $a$, $b$ and $c/sqrt{2}$ at 5.487~\AA, 5.564~\AA~and 5.500~\AA, respectively.}
    \label{Fig4}
\end{figure*}

To get better insight in the PVO film structure, we have also recorded reciprocal space maps (RSMs) with XRD experiments. In Fig.~\ref{Fig4}, we show RSMs measured from a 100~nm thick PVO films, deposited at 600$^{\circ}$C and 8.6$\times$10$^{-6}$~mbar around STO (103)$_c$ and PVO (240)$_o$ (a) , STO (013)$_c$ and PVO (33$\bar{2}$)$_o$ (b), STO ($\bar{1}$03)$_c$ and PVO (420)$_o$ (c) and STO (0$\bar{1}$3)$_c$ and PVO (33$\bar{2}$)$_o$ (d) BraggÕs peaks. We observe that PVO film is coherently grown and fully strained on the substrate from the vertical alignment of the film and substrate peaks, in good agreement with HRTEM experiments. Nevertheless, from the different reciprocal space position of (240)$_o$ and (420)$_o$ PVO peaks (indicated by an horizontal red line in Fig.~\ref{Fig3} we deduce $a\neq b$, being consistent with an orthorhombic perovskite cell. In order to refine the film lattice parameters, we have also measured RSMs around complementary ($hkl_o$) reflections : $(110)_o$, $(200)_o$, $(202)_o$, $(222)_o$ and $(204)_o$ (not shown). The refined lattices parameters are summarized in Table~\ref{Table1}. The corresponding $c$ parameter value ($c=7.809$~\AA) is twice the substrate lattice parameter $a_s$ in full agreement the HRTEM observations. We point out that the result can be only obtained by assuming a lower monoclinic symmetry ($P2_1/m$, space group 11). Indeed, PVO is also constrained along the $[1\bar{1}0]_o$ in-plane direction. Following this, the calculated $d_{1\bar{1}0}$ distance yields $d_{1\bar{1}0}=0.5\sqrt{a^2+b^2-2ab\cos\gamma}\approx3.905 \text{ \AA}=a_s$, in good agreement with experimental measurements. It therefore induces refined monoclinic angle $\gamma=88.8(4)^{\circ}$ to accommodate the substrate imposed strain. As a result, we can conclude that the PVO film epitaxy on STO imposes a lowering of the PVO structure symmetry from $Pbnm$ to $P2_1/m$. Taking into account the difference between $a$ and $b$ lattice parameters, we calculate also the angle between $[110]_o$ and $[1\bar{1}0]_o$ direction to be $\delta=2\text{tan}^{-1}(a/b)\approx89.87^{\circ}$.

In the table~\ref{Table1}, we show also the calculated strain experienced by PVO with the purpose of comparing bulk film lattice parameters. Since the angles $\gamma$ and $\delta$ a very close to 90$^{\circ}$, we have calculated the strain assuming a pseudo-orthorhombic unit cell. Hence, $\epsilon_{100}$, $\epsilon_{010}$ and $\epsilon_{001}$ are the strain acting on $a$, $b$, and $c$ lattice parameters respectively and defined as:
\begin{equation}
	\label{eq1}
	%\begin{eqnarray}
	\epsilon_{100}=\frac{a-a_0}{a_0}\text{, }
	\epsilon_{010}=\frac{b-b_0}{b_0}\text{, }
	\epsilon_{001}=\frac{c-c_0}{c_0}
	%\end{eqnarray}
\end{equation}
We determine a moderate elongated deformation along $[010]_o$ and $[001]_o$ directions ($+0.4$~\%) whereas high tensile strain acts in the PVO $[100]_o$ direction ($+1.6$~\%). We have also calculated the accumulated strain along the $[110]_o$ and $[1\bar{1}0]_o$ directions define as:
\begin{equation}
\label{eq2}
% \begin{eqnarray}
	\epsilon_{110}=\frac{d_{110}-d_{110}^0}{d_{110}^0}\text{, }
	\epsilon_{1\bar{1}0}=\frac{d_{1\bar{1}0}-d_{1\bar{1}0}^0}{d_{1\bar{1}0}^0}
%\end{eqnarray}
\end{equation}
where $d_{110}^0$ and $d_{1\bar{1}0}^0$ refer to the bulk value ($d_{110}^0=d_{1\bar{1}0}^0\approx$3.907~\AA). From the obtained strain values for the in-plane $[001]_o$ and $1\bar{1}0]_o$ directions being $+0.5$~\% and $-0.05$~\%, respectively, and that corresponding to the out-of-plane $[110]_o$ direction, we deduce a highly unbalanced strain between sample and growth direction leading to a large variation of the film unit cell volume. Thus, the elongated $[110]_o$ direction suggests an additional parameter contributing to the PVO structure distortion in addition to the epitaxial strain. 

\begin{figure*}[tp]
  \includegraphics[scale=1]{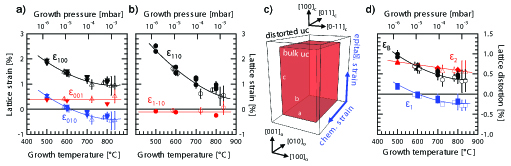}
    \caption{(a) Lattice strain as function of the growth conditions. Strain along the orthorhombic directions, $\epsilon_{100}$ (black triangle symbols), $\epsilon_{010}$ (blue triangle) and $\epsilon_{001}$ (red triangle) as a function of the growth temperature (bottom scale, full symbol) and as a function of the growth pressure (top scale, open symbols). (b) Strain along $[110]_o$ and $[1\bar{1}0]_o$, $\epsilon_{110}$ (black circle), $\epsilon_{1\bar{1}0}$ (red circle). (c) Sketch of the PVO unit cell volume change when grown on STO. (f) Representation of the distorted film structure of PVO.(d) Distortion $\epsilon_1$ (blue square),  $\epsilon_2$ (red triangle) and bulk strain (\textit{i.e.} volume expansion), $\epsilon_{B}$ (black circle symbols) dependence on growth conditions. Full symbols are for the growth temperature dependence (bottom scale) and open symbols are for the growth pressure dependence (top scale).}
    \label{Fig5}
\end{figure*}

Figure~\ref{Fig4}(a) and (b) display symmetric $\theta/2\theta$ XRD measurements for representative series of films grown at different temperatures (thickness is $\sim40$~nm) and different pressures (thickness is $\sim25$~nm), respectively, illustrating the growth conditions dependence of the PVO lattice structure. The thickness was determined using X-ray reflectometry and is confirmed by simulating the spectra with Laue fringes. Having seen that PVO films are fully strained even above 100~nm, according to our detailed HRTEM and XRD investigations and assuming that $\delta$ angle is very close to 90$^{\circ}$, we assume for all film $a$ and $b$ equal within our experimental accuracy. Consequently, we have determined the dependence of the PVO lattice parameters on growth conditions and presented in Fig.~\ref{Fig4}(c). On the one hand, the $c$-axis lattice parameter (full and open red triangles) is constant for all temperature and pressure conditions. On the other hand, we observe that $a$- and $b$-axes (full and open blue triangles) lattice parameters increase as the growth pressure decreases (5.539~\AA~and 5.569~\AA~at $1\times10^{-3}$~mbar and $8.6\times10^{-6}$~mbar, respectively), whereas $\gamma$ is reduced from $90^{\circ}$. Remarkably, the control of the growth temperature produces a similar effect on the $c$ parameter. Indeed, as the growth temperature increases (from $500^{\circ}$C up to $800^{\circ}$C) $a$ and $b$ values decreases (from 5.588~\AA~to 5.551~\AA) while $\gamma$ increases. This dependence is the consequence of a compensation of the fully strained $[110]_o$ film state, with partially out-of-plane free $a$ and $b$ axes and a $\gamma$ angle different from the orthorhombic 90$^{\circ}$, thus accommodating the highly in-plane strained state. Interestingly, we observe that while the film $a$- and $c$ lattice parameters are always above the bulk parameters, the $b-$ lattice parameter is either above or below the bulk one. Therefore, we have determined the evolution of the strains defined in Eq.~\ref{eq1} as a function of the growth conditions and displayed in Fig.~\ref{Fig5}(a). We can extract an unchanged positive $\epsilon_{001}$, a decreasing positive $\epsilon_{100}$ and a decreasing $\epsilon_{010}$ that reverses its sign at T$_{growth}\approx600^{\circ}$C and P$_{growth}\approx1\times10^{-5}$~mbar. We have also determined the strain along the $[110]_o$ and $[1\bar{1}0]_o$ following Eq.~\ref{eq2}. A negative $\epsilon_{1\bar{1}0}$ remains unchanged for all pressure and temperature conditions. However, as the pressure decreases (for a given temperature), an expansion of the PVO structure along the  $[110]_o$ direction is seen that we attribute to the creation of oxygen vacancies at low pressure~\cite{Ohtomo2007}. Furthermore, as the temperature decreases (for a given pressure) a similar expansion is observed. We explain this behavior by the known diffusion of oxygen vacancies into STO and controlled by the film/substrate oxygen exchange during the film growth\cite{Schneider2010} in out-of-equilibrium conditions (low oxygen partial pressure). Indeed the oxygen vacancy diffusion in STO is boosted during oxide thin film deposition~\cite{Herranz2009}. As the equilibrium concentration of oxygen vacancies in STO depends on the vacuum environment and also on the annealing temperature~\cite{Herranz2009}, a change of T$_{growth}$ (P$_{growth}$ is constant) would thus lead to a modification of the vacancy doping in STO, the film being a source of oxygen vacancies. Consequently, the substrate can be viewed as an oxygen reservoir for the film during its growth. Then, for this T$_{growth}$ range and at fixed P$_{growth}$, the higher the growth temperature the faster the oxygen vacancies diffuse. Based on the above consideration, we thus infer that the PVO lattice structure can be tailored by controlling the temperature and/or pressure during the deposition on STO substrate.

In the sketch of Fig.~\ref{Fig5}(c), we represent the obtained PVO unit cell when grown on PVO and we compare it with respect to the bulk one. We see that the expanded unit cell consists in the combined action of different strains acting along the $[001]_o$ and $[010]_o$ directions \textit{i.e.} epitaxial and chemical strain (referred to oxygen vacancies), respectively. Hence, we define the distortion in the $bc$-plane (being the same in the $ac$-plane) by:
\begin{equation}
	\label{eq3}
	\epsilon_{1}=\frac{1}{2}(\epsilon_{001}-\epsilon_{010})
\end{equation}
Then, it follows that the distortion ($\epsilon_2$) and the bulk distortions ($\epsilon_B$) are: 
\begin{eqnarray}
	\label{eq4}
	\epsilon_{2}=\frac{1}{4}(2\epsilon_{100}-\epsilon_{010}-\epsilon_{001})\\
	\label{eq5}
	\epsilon_{B}=\frac{1}{3}(\epsilon_{100}+\epsilon_{010}+\epsilon_{001})
\end{eqnarray} 
We note that the bulk distortion is proportional to the unit cell volume $V$ modification as $\epsilon_B=3\Delta V/V_0$ ($V_0$ being the bulk volume). In Fig.~\ref{Fig5}(d), we summarize the effect of both epitaxial and chemical strain on the PVO structure deposited on STO and controlled by the growth conditions. As the growth pressure decreases $\epsilon_B$ increases, which corresponds to a unit cell volume expansion promoted by the creation of oxygen vacancies. A similar effect occurs with decreasing the temperature $\epsilon_2$. $\epsilon_2$ increases slightly from $+0.5$~\% to $+0.8$~\%. On the contrary, $\epsilon_1$ shows that the strain changes from a tensile to a compressive state with increasing growth pressure/temperature driven by the free $b$ lattice parameter (Fig.~\ref{Fig5}(a)). 

Deposited on STO, the refined room temperature structure of PVO implies a phase structural transition from a orthorhombic to a monoclinic symmetry that otherwise occurs around 180~K in the bulk material. We anticipate that it would also influence the onset of the orbital ordering at T$_{\rm{OO}}$. Many reports have shown that RVO$_3$ orbital and spin orderings are very sensitive to structural bias~\cite{Miyasaka2003,Zhou2006}. It involves the intrinsic VO$_6$ distortion of the bulk orthorhombic perovskite structure that results in a V$-$O bond-length splitting at an octahedral site. In this study, we have shown that fine tuning of the growth conditions allows to tailor the PVO distortions. Thus, it raises the question of the magnetic response of the PVO films to the controlled structural change both from experimental and theoretical points of view.
\section {CONCLUSION}
Epitaxial PrVO$_3$ thin films have been grown on $(001)_c$-oriented SrTiO$_3$ substrate at different pressure and temperature growth conditions. By combining high resolution characterization at the meso, micro and nanoscale, we prove that bulk orthorhombic PrVO$_3$ turns into a monoclinic unit cell presenting a majoritary domain configuration consistent with a pseudo-orthorhombic $[110]_o$ unit cell with epitaxial relationships for one domain being: (i) PrVO$_3[110]_o\|$SrTiO$_3[001]_c$  and PrVO$_3[001]_o\|$SrTiO$_3[100]_c$ for one domain and (ii) PrVO$_3[110]_o\|$SrTiO$_3[001]_c$ and PrVO$_3[001]_o\|$SrTiO$_3[010]_c$ for the second domain. Moreover, after unit cell refinement and evaluation of strain values, we determine due to the out-of-equilibrium conditions, \textit{i.e} low oxygen partial pressure, and the STO-substrate/film oxygen exchange, \textit{i.e.} thermally activated oxygen vacancy diffusion in SrTiO$3$, that PrVO$_3$ films accommodate the chemical strain through the variation of the out-of-plane direction. These results open new paths alternative to substrate mismatch to control the strain and structure on PrVO$_3$ films and other similar orthorhombic perovskites by performing a fine tuning of the deposition conditions.  
\begin{acknowledgments}
The authors thank J. Lecourt for preparing high quality targets for pulsed laser deposition and F. Veillon for his valuable experimental support. Financial support by U. L\"uders through the French ANR programme GeCoDo (ANR-2011-JS080101), by the IDS-FunMat programme and by Labex EMC3 is acknowledge. P-EJ thanks the French ANR programs MOCA (2010-NANO-020-01) and NOMILOPS (ANR-11-BS10-016-02) as well as the project consortium NAO-C STProjects-133 of the FP7 project ERA.Net RUS under the grant agreement 226164. 

\end{acknowledgments}


\begin{references}

\bibitem{Imada1998}
	M. Imada A. Fujimori, and Y. Tokura,
	Rev. Mod. Phys. \textbf{70}, 1039 (1998).
	
\bibitem{Tokura2000}
	Y. Tokura and N. Nagaosa,
	Science \textbf{288}, 462 (2000).

\bibitem{Bibes2011}
     	M. Bibes, J.E. Villegas, and A. Barth\'el\'emy,
   	Adv. Phys. \textbf{60}, 5 (2011).	
	
\bibitem{Chu2008}
	Y.-H. Chu \textit{et al.},
	Nature Mater., \textbf{7}, 478 (2008).
	
\bibitem{Schlom2008}
	D.G. Schlom \textit{et al.},
	J. Am. Ceram. Soc., \textbf{91}, 2429 (2008).
	
\bibitem{Chakhalian2006}
	J. Chakhalian \textit{et al.},
	Nature. Phys. \textbf{2}, 244 (2006).
	
\bibitem{Haeni2004}
	J.H. Haeni \textit{et al.},
	Nature \textbf{430}, 758 (2004).
	
\bibitem{Rubi2009}
	D. Rubi \textit{et al.},
	Phys. Rev. B \textbf{79}, 014416 (2009).
	
\bibitem{Marti2009}
	X. Marti \textit{et al.},
	J. Magn. Magn. Mater \textbf{321}, 1719 (2009).
	
\bibitem{White2013}
	J.S. White \textit{et al.},
	Phys. Rev. Lett. \textbf{111}, 037201 (2013).
	
\bibitem{Lee2010}
	J.H. Lee \textit{et al.},
	Nature \textbf{466}, 954 (2010).
		
\bibitem{Miyasaka2003}
	S. Miyasaka \textit{et al.},
	Phys. Rev. B \textbf{68}, 100406(R) (2003).	

\bibitem{Zhou2006}
	J.-S. Zhou, and J.B. Goodenough,
	Phys. Rev. Lett. \textbf{96}, 247202 (2006).
	
\bibitem{Fujioka2005}
	J. Fujioka \textit{et al.}
	Phys. Rev. B \textbf{72}, 024460.

\bibitem{Zhou2007}
	J.-S. Zhou \textit{et al.}
	Phys. Rev. Lett \textbf{99}, 156401 (2007).
	J.-S. Zhou \textit{et al.}
	Phys. Rev. B \textbf{80}, 224422 (2009)
	
\bibitem{Bizen2008}
	D. Bizen \textit{et al.},
	Phys. Rev. B \textbf{78}, 224104 (2008).
	
\bibitem{Bizen2012}
	D. Bizen \textit{et al.},
	J. Phys. Soc. Jpn \textbf{81}, 024715 (2012).
	
\bibitem{Miyasaka2007}
	S. Miyasaka \textit{et al.},
	Phys. Rev. Lett. \textbf{99}, 217201 (2007).
	
\bibitem{Zhang}
	Q. Zhang \textit{et al.},
	arXiv:1210.6373.
	
\bibitem{Copie2013}
	O. Copie \textit{et al.}
	J. Phys: Condens. Matter \textbf{25}, 492201 (2013).

\bibitem{Mahajan1992}
	A.V. Mahajan \textit{et al.},
	Phys. Rev. B \textbf{46}, 10966 (1992)	
	
\bibitem{Onoda1996}
	M. Onoda and H. Nagasawa,
	Solid. State Commun. \textbf{99}, 487 (1996).	

\bibitem{Ren1998}
	Y. Ren \textit{et al.},
	Nature (London) \textbf{396}, 441 (1998).
		
%\bibitem{Kimishima2000}
%	Y. Kimishima \textit{et al.}
%	J. Magn. Magn. Matter. \textbf{210}, 244 (2000).
	
%\bibitem{Reehuis2006}
%	M. Reehuis \textit{et al.},
%	Phys. Rev. B \textbf{73}, 094440 (2006).
		
\bibitem{Tung2005}
	L.D. Tung,
	Phys. Rev. B \textbf{72}, 054414 (2005).			
		
\bibitem{Sage2007}
	M.H. Sage \textit{et al.},
	Phys. Rev. B \textbf{76}, 195102 (2007).	
	
%\bibitem{Prellier2000}
%	W. Prellier \textit{et al.},
%	Appl. Phys. Lett. \textbf{77}, 1023 (2000).
%	B. Mercey \textit{et al.},
%	Chem. Meter \textbf{12}, 2858 (2000).

%\bibitem{DIFFAX}
%	www.public.asu.edu/$\sim$mtreacy/DIFFaX.html
	
%\bibitem{Boullay2011}
%	P. Boullay \textit{et al.},
%	Phys. Rev. B \textbf{83}, 125403 (2011).

%\bibitem{Copie2009}
%	O. Copie \textit{et al.},
%	J. Phys.: Condens. Matter. \textbf{21}, 406001 (2009).
		
%\bibitem{Hotta2006}
%	Y. Hotta \textit{et al.},
%	Appl. Phys. Lett. \textbf{89}, 031918 (2006).

\bibitem{Ohtomo2007}
	A. Ohtomo \& H.Y. Hwang,
	J. Appl. Phys. \textbf{102}, 083704 (2007)	

\bibitem{Schneider2010}
	C.W. Schneider \textit{et al.},
	App. Phys. Lett. \textbf{97}, 192107 (2010).

\bibitem{Herranz2009}
	G. Herranz \textit{et al.},
	Appl. Phys. Lett. \textbf{94}, 012113 (2009).
						
%\bibitem{Marti2010}
%	X. Marti \textit{et al.},
%	Appl. Phys. Lett. \textbf{96}, 222505 (2010).
	
%\bibitem{Rotella2012}
%	H. Rotella \textit{et al.}
%	Phys. Rev. B \textbf{85}, 184101 (2012).
	
%\bibitem{Gajek2005}
%	M. Gajek \textit{et al.}
%	Phys. Rev. B \textbf{72}, 020406(R) (2005).
	
%\bibitem{Kumar2007}
%	A. Kumar \textit{et al.}
%	Phys. Rev. B \textbf{75}, 060101(R) (2007).
			
\end{references}
\end{document}